\documentclass{ws-p8-50}
\usepackage{emlines}
\begin{document}

\title{Phase Transition in the Random Anisotropy Model}

\author{M. Dudka}

\address{Institute for Condensed Matter Physics, Ukrainian Acad.
Sci., UA-79011 Lviv, Ukraine \\E-mail: maxdudka@icmp.lviv.ua}

\author{R. Folk}

\address{Institut f\"ur theoretische Physik,
Johannes Kepler Universit\"{a}t Linz, A-4040 Linz,
Austria\\E-mail: folk@tphys.uni-linz.ac.at}

\author{Yu. Holovatch}

\address{Institute for Condensed Matter Physics, Ukrainian Acad.
Sci., UA-79011 Lviv and
Ivan Franko National University of Lviv,
UA-79005 Lviv, Ukraine\\E-mail: hol@icmp.lviv.ua}

\maketitle

\abstracts{The influence of a local anisotropy of random
orientation on a ferromagnetic phase transition is studied for 
two cases of anisotropy axis distribution. To this end a model of a random 
anisotropy magnet is analyzed by means of the field theoretical 
renormalization group approach in two loop
approximation refined by a resummation of the asymptotic
series. The one-loop result of Aharony
indicating the absence of a second-order phase
transition for an isotropic
distribution of random anisotropy axis at space dimension $d<4$ is
corroborated. For a cubic distribution the accessible
stable fixed point leads to disordered
Ising--like critical exponents.}
%============================================================================
%                          SECTION  I
%============================================================================
\section{Introduction}\label{I}

Modern understanding of universal properties of matter in the
vicinity of critical points is mainly due to the
application of the renormalization group
 (RG) ideas.\cite{books} Applied to the problems of
condensed matter physics in the early 1970s, the
RG technique proved to be a powerful
tool to study critical phenomena. For example, expressions for
critical exponents governing the magnetic phase transition
 in regular systems are known by now with record
accuracy both for isotropic\cite{Kleinerta} ($O(m)$ symmetrical)
and cubic\cite{Kleinertb} magnets. The RG
 approach also sheds light on the influence of structural
disorder on ferromagnetism.  In the present article we will apply
the field theoretical RG approach to
study peculiarities of magnetic behaviour influenced by  disorder
in a form of random anisotropy axis.\cite{Harris73}  It is special
pleasure for us to dedicate this paper to Prof. Hagen Kleinert on
the occasion of his 60th anniversary. His contribution to the
field is hard to be overestimated.

Although an influence of a weak quenched structural disorder on
universal properties of a ferromagnetic phase
transition has already been a problem of 
intensive 
study for several decades, there remains a number of unsettled
questions. Here, one should distinguish between random site,
random-field and random anisotropy
magnets. A weak quenched disorder
preserves second-order phase transition in
three-dimensional ($d=3$) random site magnets\cite{note1} but can
destroy this transition in random field systems\cite{note2} for
$d<4$. The situation for the random-anisotropy magnets is not so 
clear.

Typical examples of random-anisotropy
magnets are amorphous rare-earth -- transition metal alloys.  Some
of these systems order magnetically and for the description
of the ordered structure it has been proposed\cite{Harris73} to
consider a regular lattice of magnetic ions, each of them being 
subjected to a local anisotropy
 of random orientation. The Hamiltonian of this random
anisotropy model (RAM) reads:\cite{Harris73}
\begin{equation}
{\cal H} =  - \sum_{{\bf R},{\bf R'}} J_{{\bf R},{\bf R'}}
\vec{S}_{\bf R} \vec{S}_{\bf R'} -D_0\sum_{{\bf R}} (\hat {x}_{\bf
R}\vec{S}_{\bf R})^{2}, \label{1}
\end{equation}
where $\vec{S}_{\bf R}$ is an $m$-component vector on a lattice
site ${\bf R}$, $J_{{\bf R},{\bf R'}}$ is an exchange interaction,
$D_0$ is an anisotropy strength, and $\hat{x}_{\bf R}$ is a unit vector
pointing in the local (quenched) random direction of an uniaxial
anisotropy.

The model has been investigated by a variety of techniques including
mean--field theory,\cite{mfa} computer simulations,\cite{comp}
$1/m$--ex\-pan\-si\-on,\cite{1overm} renormalization
group
$\varepsilon$--expansion.
\cite{Aharony75,Chen77,Pelcovits78} The limit case of an
infinite anisotropy has been subject of a detailed study as
well.\cite{infram,Fischer85}  However the nature of
low--temperature phase  in RAM is not completely clear up to now,
although several low--temperature phases were discussed
like ferromagnetic ordering,\cite{mfa,comp} spin--glass
phase,\cite{comp,1overm} and quasi long--range
ordering.\cite{qlro}

The nature of ordering is connected with the distribution of
the random variables $\hat{x}_{\bf R}$ in (\ref{1}). For
an isotropic distribution arguments similar to those applied
by Imry and Ma\cite{Imry75} for a random-field Ising model bring
about the absence of ferromagnetic order for space
dimensions $d<4$,\cite{Pelcovits78,Ma78} whereas anisotropic
distributions may lead to a ferromagnetic order.\cite{cubic}

Application of Wilson RG technique to
RAM with the isotropic distribution of a local anisotropy axis
suggests\cite{Aharony75}  the possibility of ``runaway"
solutions of the recursion equations. Such a behaviour was
interpreted as a smeared transition. However this result was
obtained in first order of 
the $\varepsilon$--expansion and
remains to be confirmed also in higher orders.

Here, we will report results obtained by means of the field
theoretical RG technique in two loop
approximation refined by a resummation of
the resulting asymptotic series. We will
consider two cases of distribution of the random anisotropy
axis and show that a ferromagnetic second-order phase
transition takes place only when the
distribution is non-isotropic. Moreover we will show that the RAM
provides another example of a disordered model, where the
only possible new critical behaviour is of ``random Ising" type,
similar to the site--diluted magnets.\cite{note1}  More
detailed results can be found in Refs.\cite{Dudka00a,Dudka00b}

%============================================================================
%                          SECTION  II
%============================================================================
\section{Isotropic case}\label{II}

In order to deal with  quenched disorder, one  way to obtain the 
effective Hamiltonian of a RAM is to make use of the replica trick. For a 
given configuration of quenched random variables $\hat{x}_{\bf R}$ in
(\ref{1}) the partition function may then be written in a form of
functional integral of a Gibbs distribution depending on
$\hat{x}_{\bf R}$. To average over
configurations one should complete the model by choosing certain
distribution of $\hat{x}_{\bf R}$. We will analyze two cases: 
first the isotropic case, where 
the random vector $\hat x$ points with equal probability in any
direction of the $m$-dimensional hyperspace, 
and second the cubic case, where $\hat x$ lies along the edges of 
$m$-dimensional hypercube (cubic case). Other distributions may be
considered as well. In the first case the distribution function
reads:
\begin{equation}
p(\hat x)\equiv\left(\int d^m\hat x\right)^{-1}
\!=\frac{\Gamma(m/2)}{2\pi^{m/2}}. \label{2}
\end{equation}
Following the above described program, one ends up with the
replica $n\rightarrow 0$ limit of the effective
Hamiltonian\cite{Aharony75}:
\begin{eqnarray}
\label{3} {\cal H}_{eff} &{=}& {-}\int d^dR \left
\{\frac{1}{2}\!\left[{\mu_0}^2|\vec{\varphi}|^2{+}
|\vec{\nabla}\vec{\varphi}|^2\right]{+}u_0 |\vec{\varphi}|^4{+}
v_0\!\sum_{\alpha=1}^n|\vec{\phi}^\alpha|^4{+}\right. \nonumber\\
&& \left. w_0\!\sum_{\alpha,\beta=1}^n\sum_{i,j=1}^{m}\!
\phi_i^\alpha\phi_j^\alpha\phi_i^\beta\phi_j^\beta \right\},
\end{eqnarray}
where ${\mu_0}^2$ and $u_0$, $v_0$, $w_0$ are defined by $D_0$ and
familiar bare couplings of an $m$-vector model, and
$\vec{\phi}^{\alpha}\equiv\vec{\phi}^{\alpha}_{\bf R}$ is a
$m$-dimensional
vector,  $|\vec{\varphi}|^2=\sum_{\alpha=1}^{n} |\vec{\phi}^\alpha|^2$.
The bare couplings are restricted by $u_0>0$, $v_0>0$, $w_0<0$. 
Furthermore, values of $u_0$ and $w_0$ are related to appropriate 
cumulants of the distribution function (\ref{2}) and their ratio 
equals $w_0/u_0=-m$. Note that the symmetry of $u_0$ and $v_0$ terms
corresponds to the random site $m$-vector model.\cite{Grinstein76}
However the $u_0$-term has an opposite sign.

In order to study long-distance properties of the Hamiltonian
(\ref{3}), we use the field-theoretical RG approach.\cite{books} Here 
the critical point of a system corresponds to a stable fixed 
point (FP) of the RG 
transformation. We apply the massive field theory renormalization
scheme\cite{Parisi80} performing renormalization at fixed space
dimension $d$ and zero external momenta.
In  two-loop approximation we get\cite{Dudka00a} expressions
for the RG functions
in form of asymtotic series in renormalized couplings
$u,v,w$.
\begin{figure} [htbp]
\begin{center}
{\input {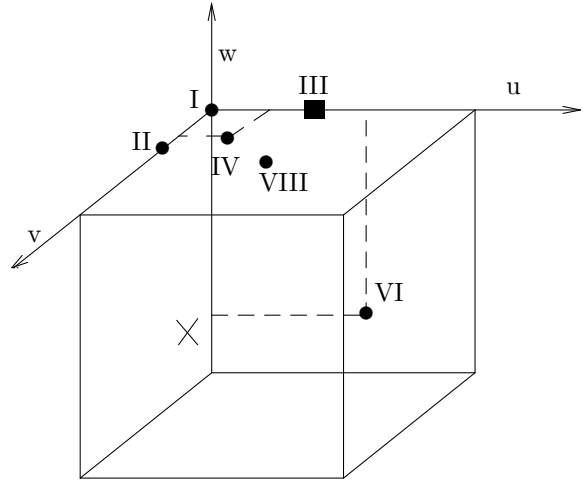}}
\end{center}
\vspace{-2.5cm} \caption{ \label{fig1} Fixed points of the RAM
with isotropic distribution of a local anisotropy axis. The 
f\/ixed points located in the octant $u>0,v>0,w<0$ are shown.  The
filled box shows the stable fixed point, the cross denotes typical
initial values of couplings.}
\end{figure}

As it was mentioned in the introduction, the only known
RG results for RAM with isotropic
distribution of the local anisotropy axis so far are those
obtained in first order in $\varepsilon$.\cite{Aharony75} In
total one gets eight fixed points.
All FPs with $u>0,v>0,w<0$ appear to be
unstable for $\varepsilon>0$ except of the ``polymer" $O(n=0)$
FP III which is stable for all $m$ (see Fig.
\ref{fig1}). However the presence of a stable FP is not a sufficient 
condition for a second-order phase transition. In 
order to be physically relevant, the FP should be 
accessible from the initial values of couplings.  This is not the case for 
the location of FPs shown in Fig. \ref{fig1}. Indeed starting
from the region of physical initial conditions (denoted
by the cross  in the Fig. \ref{fig1}) in the plane of
$v=0$ one would have to cross  the separatrix joining the
unstable FPs I and VI. This is not possible and so one never
reaches the stable FP III. As far as both FPs
I and VI are strongly unstable with respect to $v$, FP III is not 
accessible for arbitrary positive $v$ either. Finally, runaway solutions of 
the RG equations show that the second-order 
phase transition is absent in the model. 
\begin{table}[htp]
\begin{center}
\caption {\label{tab1} Resummed values of the fixed points and
critical exponents for isotropic case in
two-loop approximation for $d=3$. We absorb the value of a one-loop 
integral into  the normalization of couplings.
}
\begin{tabular}{|c|c|c|c|c|c|c|}
   \hline
FP &$m$ & $u^*$ & $v^*$ & $w^*$ &
  $\nu$ &$\eta$\\
   \hline
I&$\forall m$ &0 &0 &0& &\\ \hline II&2 &0 &0.9107 &0 &0.663
&0.027\\ &3 &0 &0.8102 &0 &0.693 &0.027\\ &4 &0 &0.7275 &0 &
0.724&0.027\\
 \hline
III&$\forall m$ &1.1857 &0 &0 &0.590 &0.023\\ \hline IV&2 &-0.0322
&0.9454 &0& 0.668&0.027\\ &3 &0.1733 &0.6460& 0&0.659 &0.027 \\ &4
&0.2867&0.4851&0&0.653 &0.028\\ \hline VI&2 &1.4650 &0 &-1.6278
&0.449 &-0.028\\ \hline VIII&2 &0.7517 &0.7072 &-0.3984  &0.626
&0.031\\ &3 &0.8031 &0.5463 & -0.3305 & 0.620&0.029\\ &4 &0.8349
&0.4545 & -0.2888 &0.617 &0.029\\ \hline
\end{tabular}
\end{center}
\end{table}
The main question of interest here is whether the above described
picture of runaway solutions is not an artifact of the
$\varepsilon$--expansion.
To check this, we used a more refined analysis of the FPs
and their stability, considering the series for
RG functions directly at
$d=3$.\cite{Parisi80} It is known that series of this type are 
at best asymptotic  and a resummation procedure
has to be applied to obtain reliable data on their
basis. We make use of Pad\'e--Borel resummation
techniques,\cite{Baker78} first writing the
RG functions as resovent
series\cite{Watson74} in one auxiliary variable and then
performing resummation. Numerical values of the
FPs are given in  Table \ref{tab1}. Resummed
two-loop results qualitatively confirm the picture obtained in first-order  
$\varepsilon$--expansion: stability of the FPs does not change after 
the resummation. This supports the conjecture of 
Aharony\cite{Aharony75} that accessible stable FP for the RAM with 
isotropic distribution of the local anisotropy axis 
 is absent. 
In the table, we list values of the 
 correlation length
and pair correlation function critical exponents $\nu$ and $\eta$ 
which are resummed in a similar way. As they are calculated in unstable FPs 
they have rather to be considered as effective ones.

%============================================================================
%                          SECTION  III
%============================================================================
\section{Cubic case}\label{III}
Let us now consider the second example of anisotropy axis
distribution, when the vector $\hat x_{\bf R}$ (\ref{1}) points
only along one of the $2m$ directions of axes $\hat k_i$ of a
cubic lattice :
\begin{equation}
p(\hat x)=\frac{1}{2m}\sum_{i=1}^m[\delta^{(m)}(\hat x-\hat k_i)+
\delta^{(m)}(\hat x+\hat k_i)]. \label{4}
\end{equation}
The {\em rationale} for such a choice  is to mimic the situation
when an amorphous magnet still ``remembers" the initial (cubic) lattice
structure. Repeating the procedure described in the previous
section, one ends up with the following effective Hamiltonian
 which is of interest in the limit $n\rightarrow
0$\cite{Aharony75}:
\begin{eqnarray}
{\cal H}_{eff}&{=}&-\int d^dR
\left\{\frac{1}{2}\left[{\mu_0}^2|\vec{\varphi}|^2{+}
|\vec{\nabla}\vec{\varphi}|^2\right]{+}u_0|\vec{\varphi}|^4{+}
v_0\sum_{\alpha=1}^n|\vec{\phi}^\alpha|^4{+}\right. \nonumber\\ &
& \left.w_0\sum_{i=1}^m\sum_{\alpha,\beta=1}^n
{\phi_i^\alpha}^2{\phi_i^\beta}^2{+}
y_0\sum_{i=1}^m\sum_{\alpha=1}^n{\phi_i^\alpha}^4\right\}.
\label{5}
\end{eqnarray}
Here, the bare couplings are $u_0>0$, $v_0>0$,
$w_0<0$. The $y_0$ term is generated  when the
RG transformation is applied
and may be of either sign. The symmetry of
$w_0$ terms differs in (\ref{3}) and (\ref{5}).  Furthermore,
values of $w_0$ and $u_0$ differ for Hamiltonians (\ref{3}) and
(\ref{5}) but their ratio equals $-m$ again.

\begin{figure} [htbp]
\begin{center}
{\input {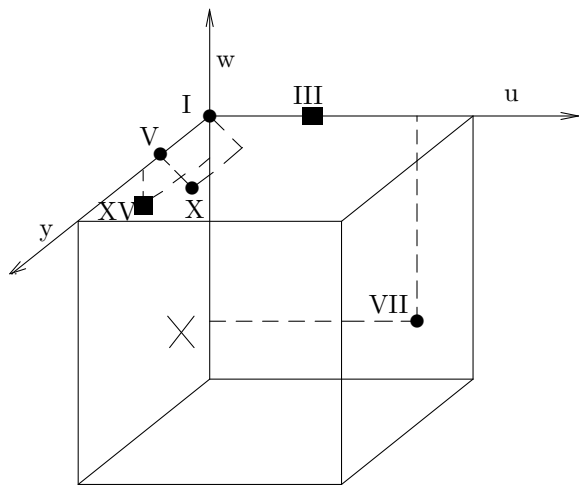}}
\end{center}
\vspace{-2.5cm}
\caption{ \label{fig2} Fixed points of the RAM with
distribution of a local anisotropy axis along hypercube axes for
$v=0$. The only f\/ixed points located in the region $u>0,w<0$ are
shown. Filled boxes show the stable fixed points, the cross denotes
typical initial values of couplings.} \end{figure}

We apply the massive field theory renormalization
scheme\cite{Parisi80} and get the RG
functions in two-loop approximation.\cite{Dudka00b} As  in
the previous case we reproduce the first-order
$\varepsilon$-results.\cite{Aharony75} Now one gets 14 FPs.
However, in first order of the $\varepsilon$--expansion all FPs
with $u>0,v>0,w<0$ appear to be unstable for $\varepsilon>0$
except of the ``polymer" $O(n=0)$ FP III which
is stable for all $m$ but not accessible (see Fig. \ref{fig2}). Now
the account of the $\varepsilon^2$-terms qualitatively changes the
picture. Indeed, the system of equations for the FPs appears
to be degenerated  at the one loop level. As known from
other cases in two loop order this leads to the appearance of a
new FP which is stable and is expressed by a
$\sqrt{\varepsilon}$ series.\cite{Grinstein76} The
possibility of such scenario was predicted already in
Refs.\cite{cubic} However it remained unclear whether there exist
any other accessible stable FPs.

Applying  Pad\'e--Borel resummation we get
16 FPs. Values of the FPs with $u^*>0,v^*>0,w^*<0$ are
listed in the Table \ref{tab2}. The last FP XV in Table 
\ref{tab2} corresponds to the stable FP of 
$\sqrt\varepsilon$-expansion. It has coordinates with $u^*=v^*=0$, $w^*<0$ 
and $y^*>0$ and is accessible from the typical initial values of couplings 
(shown by a cross in the Fig. \ref{fig2}). Applying the 
resummation procedure, we have not found any other stable 
FPs in the region of interest. The effective Hamiltonian (\ref{5}) at 
$u=v=0$ in the replica limit $n\rightarrow0$ reduces to a product of
$m$ effective Hamiltonians of a weakly diluted quenched random
site Ising model. This means that for any value of $m>1$ the system
is characterized by the same set of critical
exponents as those of a weakly
diluted quenched random site Ising model. In the Table, we give 
\ref{tab2} values of critical exponents in
the other FPs as well: if the flows from initial values of couplings
pass near the these FPs, one may observe an effective critical
behaviour governed by these critical exponents.

\begin{table}[htbp]
\begin{center}
\caption
{\label{tab2} Resummed values of the FPs and critical
exponents for cubic distribution
in two-loop approximation for $d=3$. We absorb the value of a one-loop 
integral into  the normalization of couplings.
}
   \begin{tabular}{|c|c|c|c|c|c|c|c|}
   \hline
FP &$m$ & $u^*$ & $v^*$ & $w^*$ & $y^*$ &  $\nu$ &$\eta$\\
   \hline
   I&$\forall m$ &0 &0 &0&0 &1/2 &0 \\
   \hline
 &2 &0 &0.9107 &0& 0 &0.663 &0.027\\
II&3 &0 &0.8102 &0 & 0&0.693 &0.027\\
&4 &0 &0.7275 &0 & 0&0.720&0.026\\ \hline
 III&$\forall m$ &1.1857 &0 &0& 0 &0.590 &0.023 \\
 \hline
V&$\forall m$ &0 &0 &0 &1.0339&0.628 &0.026\\ \hline
 VI&3 &0.1733
&0.6460 &0 & 0&  0.659&0.027\\
&4 &0.2867 &0.4851 & 0& 0&0.653 &0.027\\
\hline
 VII&$\forall m$ &2.1112 &0 &-2.1112 &0&1/2 &0\\ \hline
 &2 &0 &1.5508 &0 &-1.0339& 0.628 &0.026\\
 VIII&3 &0 &0.8393 &0 &-0.0485& 0.693 &0.027\\
 &4 &0 &0.5259 &0 &0.3624& 0.709 &0.026\\
\hline
 IX&3 &0.1695 &0.7096 & 0& -0.1022&0.659& 0.027\\
 &4 &0.2751 &0.4190 & 0& 0.1432&0.653 &0.027\\
\hline
X&$\forall m$ &0.6678 &0 &-0.6678 &1.0339&0.628 &0.026\\
\hline
XV&$\forall m$ &0&0&-0.4401&1.5933&0.676&0.031\\
\hline
\end{tabular}
\end{center}
\end{table}

%============================================================================
%                          SECTION IV
%============================================================================
\section{Conclusions}\label{IV}
We applied the field theoretical
RG approach to  analyze critical
behaviour of a model of random anisotropy
magnets with isotropic and cubic distributions of a local
anisotropy axis. The origin of a low--temperature phase in this
model is not completely clear. General arguments based on an
estimate of the energy for formation of magnetic
domains\cite{Imry75} lead to the conclusion that for $d<4$ a
ferromagnetic order is absent.\cite{Pelcovits78,Ma78} However,
these arguments do not take into account entropy which may be
important for disordered systems.\cite{Fischer85} Furthermore,
they do not apply  for anisotropic distributions
of the random axis.\cite{cubic}

In the RG analysis the absence of a
ferromagnetic second-order phase transition corresponds to  the 
lack of a stable FP of the RG
transformation. However in the case of RAM with isotropic
distribution of a local anisotropy axis the scenario differs. Our
two-loop calculation leads to a $O(n=0)$
symmetric FP which is stable for any value of
$m$ for both isotropic and cubic distributions of a random
anisotropy axis. Note that this
FP is not accessible from the initial values of
the couplings. We checked the location of the FPs up to
second order in the 
$\varepsilon$--expansion and by
means of a fixed $d=3$ technique refined by Pad\'e--Borel
resummation.

In the case of isotropic distribution of a random
anisotropy axis our analysis supports
the conjecture of Aharony\cite{Aharony75} based on results linear in
$\varepsilon$ about runaway solutions of the
RG equations. For the cubic
distribution we get two stable FPs. One of them (FP III in Fig. 
\ref{fig2}) is not accessible as in the isotropic case.  But the disordered 
Ising-like FP (FP XV in Fig.  \ref{fig2}) may be reached 
from the initial values of couplings. Applying the resummation procedure we 
have not found any other stable FPs in the region of interest. This
means that RAM with cubic distributions of a random
anisotropy axis is governed by a  set of
critical exponents of a weakly site
diluted quenched Ising model.\cite{Grinstein76,note4}

To conclude we want to attract attention to a certain similarity in
the critical behaviour of both random-site\cite{Grinstein76} and
random-anisotropy\cite{Harris73} quenched magnets:
if there appears a {\em new critical behaviour} it {\em always is
governed by critical exponents of
site-diluted Ising type}. The above calculations of a ``phase diagram"
of RAM are based on two-loop expansions improved by a resummation
technique. Once the qualitative picture
 becomes clear there is no need
to go into higher orders of a perturbation theory as far as the
critical exponents of site-diluted Ising model are known by now with
high accuracy.\cite{note4}

%============================================================================
%                          Acknowledgement
%============================================================================
\section*{Acknowledgments}
Yu. H. acknowledges helpful discussions with  Mykola Shpot. This work
has been supported in part by "\"{O}ster\-rei\-chi\-sche Nationalbank
Ju\-bi\-l\"a\-ums\-fonds" through the grant No 7694.

\end{document}